\documentclass[review]{elsarticle}
\usepackage{lineno,hyperref}
\usepackage{relsize}
\usepackage{amsmath}
\journal{Journal of \LaTeX\ Templates}

\begin{document}

\begin{frontmatter}

\title{New fiber read-out design for the large area scintillator detectors: providing good amplitude and time resolutions}

\author{V. Grabski\fnref{myfootnote}}
\address{Instituto de Fisica Universidad Nacional Autonoma de Mexico, Mexico}
%%\fntext[myfootnote]{grabski@fisica.unam.mx.}

\cortext[mycorrespondingauthor]{Corresponding author}
\ead{varlen.grabski@cern.ch}

\address[mymainaddress]{Av Universidad 3000, Instituto de Fisica, Coyoacan, Mexico}

\begin{abstract}
Most of the time-of-flight systems as well as the fast interaction trigger detectors have large surfaces and the principal requirements for the above mentioned detectors is a good time resolution of the order $100-200ps$. The easiest solution here is to split the large surface into the small tiles with a read-out directly attached to the photo-sensor. This solution is expensive because the number of the channels grows proportional to the surface. Although if the coverage of the whole surface by the sensors is not sufficient, one can obtain non uniform response, which is not acceptable if uniformity is required. Our suggestion is based on the usage of clear fibres mounted perpendicular on the surface of the scintillator as a matrix providing uniform surface reduction and keeping most of the benefices of the  small tiles with the photo-sensors. In this work the design and some analytic estimations of the light collection  efficiency for direct and reflected photons will be presented. Also some simple estimations are presented for the time-spread and for the light pulse wave form dependent on the lateral sizes and the thickness of the scintillator.
\end{abstract}

\begin{keyword}
\texttt{Scintillation detector, Time-of-flight detector, Particle identification, Fast Interaction Trigger, Optical Fibre}
\end{keyword}

\end{frontmatter}

%%\linenumbers

\section{Introduction}
Introduction of WLS fibres resolve many problems of the light collection like uniformity, compact and flexible light transport. These light collection systems were used mainly in calorimetry and other applications where the timing characteristics are not important \cite{ATLAS}. After introduction of fast WLS fibres becomes possible to use them also for the applications when good time resolution(~1ns) is required \cite{V0-perf}. Anyway the use of WLS will include additional stochastic process, that spreads in the time of fast photons from scintillator and worsen time resolution.
 Usually standard light guide designs as well as WLS systems using a single photo-sensor includes time spread due to coordinate spread related lateral sizes of the scintillation tile. To avoid this, normally  small scintillator tiles with the direct coupling photo-sensors to the scintillator surface have been used \cite{Luk}.  This design is good for timing, but can introduce non uniform surface response if photo-sensor surface is considerably smaller than the scintillator surface. On the other hand this design will increase the number of photo-sensors as well as the number of electronic channels. In this work the proposed design  provides a uniform reduction of the scintillator surface up to the photo-sensor surface and conserves the fast photons timing as in the case of direct coupling. Of course using long fibres will spread photons in the time, because of the angular aperture of the fibre transmission and will worsen time characteristics. Here by the fibre density variation one can reach maximum surface reduction to achieve the required time characteristics. Sometimes it is important when the large dynamic range for the amplitude measurement is required. This design also provides a volume uniformity response, which is very important for the detectors with large volumes and when the pinpoint light pulse should be detected. The new design is based on the usage of clear fibres having the same longitude and are mounted perpendicularly to the large surface of the scintillator. The usage of other types of surface reducers like concentric cone or fibre optic taper requires large space for the detector and do not provide uniform surface response.The principal disadvantages of this design is the large number of fibres that is proportional to the surface and can be of order of few tenths of thousand per square meter. Historically this scheme of the light collection was proposed for the large area scintillator detector V0+ of FIT(Fast interaction trigger)\cite{Wladek}   for the ALICE experiment\cite{ALICE-up}. This scheme have been intensively studied  for a few years and demonstrated that can be used for the V0+ detector, which is already constructed and will be installed at the end of 2020.
 In the chapter of "Fibre read-out design" it will be shown basic light collection properties, which can be generalized for the WLS design as well. In the chapter of the "Light collection efficiency estimation"  will be presented some analytic relations for the direct photons from the light source and for some well defined reflectors like mirror and Lambert diffuse reflection. In Appendixes  some calculations for the light collection efficiency considering different reflection conditions are presented.

\section{Fibre read-out design}

Schematic view of the design is shown in Fig 1. The fibres are mounted perpendicularly to the scintillator surface like a matrix. For the simplicity, here we demonstrate a square tiling, but one can use any type of tiling dependent on the detection task.  This tiling has Cartesian coordinate symmetry and the distance ($d_{f}$) between neighbour fibres will define the fibre density ($D_{f}$) on the unit surface as $D_{f} = 1/d_{f}^{2}$. There are three principal ways to couple fibres with the scintillator: -a direct one with an optical contact (see Fig 1(a)); -non direct with small light concentrators for each fibre shown in Fig 1(b) and between the scintillator and fibre matrix there is a plastic layer Fig 1(c). The design in Fig 1(a) is applicable if the scintillator thickness is much larger than the fibre distance. For this design if the light source is close to the fibres(when one or a few fibre are participating in the light collection) it will provide relative large light collection efficiency that will affect surface uniformity. Also there are shadow regions between fibres with very low light collection efficiency, because of limited fibre transmission angle.  The design shown in Fig 1(b) provides better surface uniformity reducing small distance effects of the light source. Although this design is preferable, but is more complex for the construction. Also this design has no shadow regions like the design in Fig 1(a) which became important if scintillator thickness is comparable with the fibre distance. The shadow regions as well as the small distance effects also can be reduced using the design shown in Fig 1(c) that is easier for the construction than the one with light concentrators and provides better uniformity than direct coupling of fibres. In this case the thickness of the plastic shown in Fig 1(c)can be estimated by the fibre distance to eliminate shadow regions. All three options of mounting fibres have similar characteristics and the difference between them is due to small differences in uniformity and small distance effects. With these small differences it can be shown(see the next chapter) that this design in general provides volume uniform light collection efficiency.  

\section{Light collection efficiency estimation}

For the analytic estimation of the light collection efficiency we introduce a few definitions as: the fibre is an object that will transmit  the light within some angular acceptance $\Theta$(which depends only on the fibre type) with an efficiency $\varepsilon_{f}$ (which depends on the fibre type, longitude, light wave length and bending radius); between the fibre and the scintillator there is an optical contact that is transmitting the light with the efficiency $\varepsilon_{sf}$ and between the fibre and the photo-sensor the transmission efficiency is $\varepsilon_{fp}$. So for a given point of the light source inside the scintillator volume the light collection scheme for the direct photons (photons from the source without reflections from the scintillator borders) $DP$ can be drown as it is shown in Fig 2.
 As it can be seen from Fig 2 for the $DP$  the number of  the fibres $N_{f}$,which are participating in the light collection can be estimated as a:
 \begin{equation}
 N_{f}=D_{f} \times\pi \times h^{2} \times  \tan^{2}\Theta
 \end{equation}
where ($h$) is the distance of the light source  from the plane of the fibre matrix. For this point light source a given fibre with the index $i$ will have approximately a solid angle acceptance:
\begin{equation}
\Delta\Omega_{i} \approx \frac{S_{f} \times cos^{3}\theta_{i}}{h^{2}}
\end{equation}

where $S_{f}$ is the fibre surface and $\theta_{i}$ is the angle between the vector connecting the light source to the given fibre center  and the vector that is perpendicular to plane of the fibre matrix(see Fig 2). So for the number of photons that can be transmitted up to photo-sensor by the given fibre $N_{i}(h)dh$ from the light source(assuming uniform light emission) with the size $dh$ can be written:

\begin{equation}
N_{i}(h)dh = \frac{N_{0} \times F_{tr} \times S_{f} \times \exp(-\frac{a(h,l)}{cos\theta_{i}}) \times cos^{3}\theta_{i}}{4\times \pi \times h^{2} \times H}dh
\end{equation}

where $N_{0}$ is the number of photons emitted in the whole scintillator thickness of $H$ with the uniform emission in $4\times\pi Sr$ solid angle, $a(h,l) = h/\lambda_{sc}+l/\lambda_{f}$ is the distance of the light source  from the photo-sensor in attenuation length units(where $\lambda_{sc}$ and $\lambda_{f}$ are attenuation lengths of the scintillator and the fibre correspondingly, l is the fibre length), $F_{tr}= \varepsilon_{f}\times \varepsilon_{sf}\times\varepsilon_{fp}$ is the transmission efficiency through the surfaces scintillator-fibre and fibre-photo-sensor. Here we assume the approximation that the transmission through the surfaces is almost independent from the angle for the small angles  $\theta \leq 30^{o}$ and from the position of light emission.
 Then the number of all photons($N_{p}(h)dh$) from a given point of the light source that will be transmitted up to the photo-sensor can be estimated as a sum for all $N_{f}$ fibres assuming that the transmission properties are the same for all fibres:
 
\begin{eqnarray}
\nonumber N_{p}(h)dh = \frac{N_{0} \times F_{tr} \times S_{f}\times N_{f} }{4\times\pi\times h^{2}\times H} \\
\times\lbrace\frac{1}{N_{f}}\times {\mathlarger{‎‎\sum}}_{i=0}^{N_{f}‎} \exp(-\frac{a(h,l)}{\cos\theta_{i}}) \times \cos^{3}\theta_{i}\rbrace dh  
\end{eqnarray}

Combining Eq. (1) and (4) and integrating for the scintillator thickness for the total number of the collected direct photons $N_{DP}$ can be written:

\begin{eqnarray}
\nonumber N_{DP} = \int_{0}^{H} N_{p}(h)dh = \frac{N_{0} \times F_{tr} \times S_{f}\times D_{f}\times \tan^{2}\Theta  }{4\times H} \\
\times \int_{0}^{H}\left( \frac{1}{N_{f}} \times {\mathlarger{‎‎\sum}}_{i=0}^{N_{f}‎} \exp(-\frac{a(h,l)}{\cos\theta_{i}}) \times \cos^{3}\theta_{i} \right) dh  
\end{eqnarray}

The dependence of $N_{DP}$ from the distance $h$ that is included in $a(h,l)$  is as $h/\lambda_{sc}$. If $h\ll\lambda_{sc}$ then $N_{DP}$ is approximately  independent from the distance of the matrix plane. Of course the condition of $h \geq d_{f}$ will be fulfilled having at least few fibres involved in the light collection.   Most of the fast scintillators have attenuation lengths larger than 100cm, so this approximation could be good enough for the scintillator thickness up to 10cm. For the designs shown in Fig 1(b and c) the condition of $h \geq d_{f}$ is already fulfilled so it can be achieved better volume uniformity for entire scintillator volume. 
The calculation of the integral in (5) to estimate of $N_{DP}$ is performed in Appendix A. Obtained analytic expression is a large one, that’s why we present here as a function $\Phi_{dp}(\cos\Theta,a_{0},a_{1})$:

\begin{eqnarray}
\nonumber N_{DP} = \frac{N_{0} \times F_{tr} \times S_{f}\times D_{f}\times \tan^{2}\Theta  }{4\times H} \times \Phi_{dp}(\cos\Theta,a_{0},a_{1}) 
\end{eqnarray}

where $a_{0}=a(0,l)$ is the fibre longitude in attenuation longitude units and $a_{1}=a(H,l)$ is the sum of the scintillator thickness and the fibre longitude in attenuation longitude units.
 The light collection efficiency for $DP$ can be estimated as $\Upsilon_{DP}(\cos\Theta,a_{0},a_{1})$ = $N_{DP}$/$N_{0}$. 
 
\begin{equation}
\Upsilon_{DP} = \frac{N_{0} \times F_{tr} \times S_{f}\times D_{f}\times \tan^{2}\Theta  }{4\times H} \times \Phi_{dp}(\cos\Theta,a_{0},a_{1}) 
\end{equation},
which converts a relation independent from $H$ if we ignore attenuation in the scintillator:
\begin{equation}
\Upsilon_{DP} = \frac{N_{0} \times F_{tr} \times S_{f}\times D_{f}\times \tan^{2}\Theta  }{4} \times \Phi_{dp}(\cos\Theta,a_{0},a_{0}) 
\end{equation}
 
Ignoring also light attenuation in the fibres we obtain a simple expression. 
 
\begin{equation}
\Upsilon_{DP} = \frac{F_{tr} \times S_{f}\times D_{f}\times \tan^{2}\Theta(1-\cos^{4}\Theta)}{16(1-\cos\Theta)}
\end{equation}

 The simplified relation of the light collection efficiency for the DP depends on the three parameters like the fibre surface $S_{f}$, the fibre aperture angle $\Theta$ and the fibre density $D_{f}$. One can easily  increase or decrease the efficiency manipulating with these parameters. Direct photons are only a part of the light that can be collected. Painting all scintillator borders with the black color one can reject the contribution from the reflected photons. The light collection efficiency will increase considering the light reflection from the scintillator borders. 
 For example in case of specular reflecting coating on the large surface of the scintillator in front of the fibre matrix will increase the light collection efficiency approximately twice due to the solid angle increase (see Appendix A). If matrix surface also has specular reflective coating then the light collection efficiency can be estimated as a sum of all reflections. The number of the reflection is practically  limited, because of the efficiency reflection from the surfaces, which is normally smaller than 0.8 for blue scintillator light. So $10$ reflections are sufficient to estimate efficiency within a few percent precision.
 
   In case of the usage of the diffuse reflector coating for the scintillator the light collection efficiency also depends  on the lateral sizes, because in this case large areas of the surface can have contribution in the light collection. Of course everything depends on the geometry and for the estimation efficiency in general we will perform it for the infinite lateral sizes of the scintillator, because in that case it is easy to obtain analytic expressions. For this estimation we should consider some real construction design, for example, the one that was used for the large scintillator area of the ALICE FIT detector\cite{varlen}.
The simple light collection drawing for the above mentioned detector for the infinite lateral sizes is shown in Fig 3. As it can be seen from the figure the whole surface against the fibre matrix as well as the fibre matrix surface participate in the light collection processes   if they both have diffuse reflection properties. Here we are going to consider it a Lambert type reflection (the reflected light has $cos\theta$ low from the zenith angle) because in this case it is easier to obtain analytic estimations. As it can be seen from Fig 3 for each small solid angle, emission from the point source after reflection can contribute in the light collection process. So for the infinite lateral sizes of the scintillator for the first reflection only the half of the light emission can participate in the light collection. The other half will contribute for the multiple reflections, if the matrix surface also has diffuse reflective coating. The number of significant reflections is limited within a few ns  after more than 10 reflections the emitted light practically will be absorbed or lost because of the average reflection efficiency, which is below 80\%(depends on the emission spectrum).  So for the first reflection and for a given fibre $i$ using similar procedure like (3-6) for the number of transmitted photon $N^{i}_{df1}(\theta_{1},\theta_{2i},h_{1},H)$ assuming a Lambert type of reflection, can be written:

\begin{eqnarray}
\nonumber N^{i}_{df1}(\theta_{1},\theta_{2i},h_{1},H) d\Omega_{1}dh_{1} = \frac{N_{0}}{4\times\pi\times H} \times \exp(-\frac{b(h_{1})}{\cos\theta_{1}})d\Omega_{1} dh_{1} \\ 
 \times\left( \frac{F_{tr}\times \epsilon_{dfr}}{\pi} \times\exp(-\frac{a_{1}}{\cos\theta_{2i}})\times \cos\theta_{2i}\times d\Omega_{2i}\right),
\end{eqnarray}

where the expression before the parentheses is the number of photons reaching the reflection surface within the solid angle $d\Omega_{1}=d\phi\times d\cos\theta_{1}$, $\phi$ is the azimuthal angle,  $\theta_{1}$ is the zenith angle (see Fig 3), expression in the parentheses is the portion of reflected light that is transmitted up to phtosensor by the fibre $i$, $h_{1}$ is the distance of the light source from the reflection plane, $b(h_{1}) = h_{1}/\lambda_{sc}$, $a_{1}$ is the same as in Eq. (6), $\epsilon_{dfr}$ is the average efficiency of the reflection of the diffuse coating and $\theta_{2i}$ is the angle between the vector connecting the light source on the reflection plane to a given fibre center and the to vector that is perpendicular to the fibre matrix plane(see Fig 3), $d\Omega_{2i}$ is the solid angle of the fibre $i$ from the light source located on the reflective surface and approximately can be estimated using expression (2).
The total number of the transmitted photons $N_{df1}(\Psi,H)$ can be estimated summing for the all participating fibres $N_{fH}$, replacing value of $N_{fH}$ using expression (1) and performing the integration of: $d\Omega_{1}$ by $\theta_{1}$ and $\phi$ ($\theta_{1}$ from zero to $\Psi$ and $\phi$ from zero to $\pi$) and by $h$(from zero to $H$):

\begin{eqnarray}
\nonumber N_{df1}(\Psi,H) = {\mathlarger{‎‎\sum}}_{i=0}^{N_{fH}‎}N^{i}_{df1} =\frac{N_{0} \times \epsilon_{dfr}\times F_{tr} \times S_{f}\times  D_{f} \times\tan^{2}\Theta}{4\times H} \\ 
\times \int_{0}^{H}\int_{\Psi}^{0} \exp(-\frac{b(h_{1})}{\cos\theta_{1}})d\cos\theta_{1} dh_{1} \times\left\lbrace \frac{1}{N_{fH}} \times {\mathlarger{‎‎\sum}}_{i=0}^{N_{fH}‎} \exp(-\frac{a_{1}}{\cos\theta_{2i}}) \times \cos^{4}\theta_{2i} \right\rbrace
\end{eqnarray}

After integration and sum estimation in the expression (10), which is obtained in  Appendix B, the result is shown below:

\begin{eqnarray}
\nonumber N_{df1}(\Psi,H) = \frac{N_{0} \times \epsilon_{dfr}\times F_{tr} \times S_{f}\times  D_{f} \times\tan^{2}\Theta  }{4\times H} \\
\times I(\Psi,H)\times S_{dfr}(a_{1},\Theta) 
\end{eqnarray}
   
where $I(\Psi,H)$ is the integral in relation (9) that is the part of the emitted light reaching the reflection surface and $S_{dfr}(a_{1},\Theta)$ is the value of the sum in the parenthesis in the expression (10). For the infinite lateral sizes $\Psi = 90^{o}$

If the plane of the fibre matrix plastic support also has a diffuse reflectivity then we should sum all of reflections to obtain total light collection efficiency. For the matrix plane the reflection efficiency should be reduced by the factor of $\varepsilon_{fm} = 1-S_{fib}/S_{tot}$, which is connected with the light loose in fibres (see Appendix A).  So for the number of transmitted photons of a double reflection (at the beginning from the matrix plane and then from the plane in front of the fibres)  $N_{df2}(\Psi,H)$, can be obtained an expression in similar way:
  
\begin{eqnarray}
\nonumber N_{df2}(\Psi,H) = \frac{N_{0} \times \epsilon_{dfr}\times F_{tr} \times S_{f}\times  D_{f} \times\tan^{2}\Theta  }{4\times H} \\
\times \epsilon_{dfr}\times \varepsilon_{fm} \times\varepsilon_{sfs} \times I(\Psi,H)\times S_{dfr}(a_{2},\Theta) 
\end{eqnarray}  
  
where $a_{2}=a(2H,l)$ and  $\varepsilon_{sfs}$ is the efficiency of the light transmission between the scintillator and the fibre plastic support.Here we ignore the light attenuation in the thin layer of the fibre plastic support, which is much smaller than in the scintillator. So in this way we can estimate any time of the reflections just modifying parameter $a$ and multiplying average reflection efficiency for the surface.  Number of reflections is limited because of the value of $\varepsilon_{dfr} \leq 0.8$ as well as $\varepsilon_{fm} \leq 1.$.

  The light collection efficiency with diffuse reflection for $n$ reflections and for infinite lateral sizes can be estimated as:

\begin{eqnarray}
\nonumber \Upsilon_{DFR} = \frac{ \epsilon_{dfr}\times F_{tr} \times S_{f}\times D_{f}\times\tan^{2}\Theta  }{4 H}\times  I(\Psi,H)\times
(S_{dfr}(a_{1},\Theta) +\\+\epsilon^{2}_{dfr}\times \varepsilon^{3}_{sfs}\times \varepsilon_{fm} \times S_{dfr}(a_{2},\Theta)+...+ \epsilon^{2n-2}_{dfr}\times \varepsilon^{2n-1}_{sfs}\times \varepsilon^{n-1}_{fm} \times S_{dfr}(a_{n},\Theta)) 
\end{eqnarray} 

where $a_{n}=a(nH,l)$. In case of short fibre lengths the attenuation can be ignored and for the light collection efficiency can be obtained a simple expression for the first $n$ reflections that is shown below(see Appendix B):

\begin{eqnarray}
\nonumber \Upsilon_{DFR} =  \epsilon_{dfr}\times F_{tr} \times S_{f}\times D_{f} \times\tan^{2}\Theta  \times(1-\cos\Psi)\times\\
\frac{(1-cos^{5}\Theta)}{10\times(1-cos^{2}\Theta)}(1+\epsilon^{2}_{dfr}\times \varepsilon^{3}_{sfs}\times \varepsilon_{fm} +...+ \epsilon^{2n-2}_{dfr}\times \varepsilon^{2n-1}_{sfs}\times \varepsilon^{n-1}_{fm}) 
\end{eqnarray} 

For the average reflection efficiency about $\epsilon_{dfr}=0.8$ up to 7-8 reflection will give significant contributions.  So using diffuse reflective coating the amount of the light can be increased up to 4 times compared with the direct photons. So for the tasks when the amplitude resolution is important, the usage of diffuse reflection is preferable. If the detector lateral sizes are limited, which is usual case in the real life, it is also possible to perform analytic estimations considering reflections from the lateral planes. This can be performed in a similar way and it should be done for the already defined geometry. The total light collection efficiency $\Upsilon_{TOT}$ can be estimated as a sum of the direct and the reflected light collection efficiencies:

\begin{eqnarray}
\nonumber \Upsilon_{TOT} =  \Upsilon_{DF}+\Upsilon_{DFR} 
\end{eqnarray} 

All analytic expressions except when the attenuation is ignored include the Exponential Integral, which can be evaluated only numerically. For this reason we make two plots to show the dependencies of the light collection efficiency from the number of reflections and from the lateral size. Here we ignore the reflections from the lateral sides to simplify the calculations. For these calculations the following parameter values have been used($\lambda_{sc}=100cm$,$\lambda_{f}=200cm$,$H=4cm$,$l=50cm$,$\Theta=21$,$S_{f}=0.01\pi/4 cm^{2}$,$D_{f}=4/cm^{2}$,
$F_{tr} = 0.77$,$\epsilon_{dfr} = 0.8$), which are somehow arbitrary, but not so far from the reality. Results are shown in Fig. 4 and Fig.5. As it can be seen from the Fig.4 for the used parameters 6-8 reflection are sufficient to estimate light collection efficiency. In fig.5 is shown the dependence of light collection efficiency from the lateral sizes in scintillator thickness units. As it can be seen from the figure the efficiency will increase significantly for the lateral sizes up to 8 scintillator thickness and then almost is saturated. For real detectors lateral sides also contribute in reflections so we will not have this fast increase for small lateral sizes. Anyway for this design the light collection efficiency depends on the lateral sizes. For the small lateral sizes one can note that the total efficiency is factor of 3 is larger than the one with only direct photons. This already have been observed for a 6x6x3cm3 small prototype covered with the black paper and with the white paint. Of course for the cover with the black paper there will some internal reflections from scintillator surfaces, but the factor is similar to the value that  predicts  calculations. 

\section{Time spread estimation}
The time spread for $DP$ photons that is important for the fast timing has three principal sources: -the stochastic process of light emission; -the fibre longitude because of the transmission light angle aperture; -the thickness of the scintillator. The main time spread comes from the stochastic light emission process of the scintillation material.
To have good time resolution it is required that the light collection system should have sufficient efficiency and less time spread for the photon transport. 
 For DP the estimation of the time spread due to photon transport is simple and is done in Appendix C using the light path spread due to angle and photon position variation inside the scintillator. The obtained relation for the relative value of time spread(standard deviation over to average value) assuming that the light speed is the same in the scintillator and in the fibre as it is shown below:

\begin{equation}
\frac{\sigma_{t}}{\overline{t}} =  \sqrt{\frac{(1+\frac{H}{l})\times (\cos\Theta-1)^{2}} {((1+\frac{H}{2l})^{2}\times\cos\Theta\times\ln^{2}\cos\Theta}-1}  
\end{equation}

where $\Theta$ is the value of the fibre aperture angle, $\overline{t}$ is the average value of the time transition to the photo-sensor, $l$ is the fibre length and $H$ is the scintillator thickness. As it can be seen from the formula() the main time spread comes from the fibre length if the  thickness of the scintillator is much smaller than the fibre longitude. In this case the time spread is proportional to average transition time which in its turn is proportional to the fibre longitude. For the single clad fibre with NA(numerical aperture) ~ 0.5 the coefficient is about 0.02. For the multi-clad fibres this coefficient will increase up to 0.03(using formula C6). So this means that for the time performance is better to use higher density of the fibres and a single clad than a multi-clad fibres. Of course this estimation should be considered as a limit that can be achieved for the given $H$ and $l$.

For the specular reflection from the large surface of the scintillator the thickness of the scintillator will be doubled. In case of the diffuse reflection the time spread estimation is not as easy as in the previous case. To study the time spread for the diffuse reflection a simple simulation geometry  have been used (shown in Fig 6). All important simulation parameters are mentioned in the previous chapter. 
 The simulation results for the pulse time wave forms of direct and reflected photons with the large statistics just to have smooth curve is shown in Fig. 6. As it can be seen from the figure the number of the reflected photons is significantly larger(factor of 3) than DP and the rise time of fast photons is less than $100ps$. The delay of the twice reflected photons is about $400ps$, which corresponds to the thickness of scintillator. The standard deviation of the time spread of $DP$ is about 112 ps, which is similar to the approximate value obtained from formula (C6).  In the same figure is also shown the modification of the time wave forms when the scintillator stochastic time emission is included. As it can be seen from the figure the rise time of the light pulse mainly depends on $DP$ and the once reflected photons. So if the photon statistics and the surface uniformity is not important the multiple reflections can be excluded painting the fibre plane surface with the black color. 

\section{Conclusions}

In this study is presented a new light collection design using clear fibres and some analytical expressions to estimate expected light collection efficiency.  Historically this scheme was proposed for the FIT(Fast interaction trigger) large area scintillator detector V0+  of  the ALICE experiment.  The obtained approximate analytical expressions will be useful to understand  the light collection scheme as well as for the development of a real detectors in general. Experimental studies for V0+ prototypes a qualitative agreement between some of the results of these calculations have been already observed.

\section{Acknowledgements}
Author acknowledge the partial support from PAPIIT-UNAM  IN111117, and CONACYT  280362 grants.

\appendix \label{Appendix}
\section{Efficiency estimation for direct photons}\label{Efficiency estimation for direct photons}

Before estimating the integral in the expression (5) or $\Phi_{dp}(\cos\Theta,a_{0},a_{1})$ first of all it should be estimated the sum shown below:
\begin{equation}
S(a,\Theta) =  \frac{1}{N_{f}} \times {\mathlarger{‎‎\sum}}_{i=0}^{N_{f}‎} \exp(-\frac{a(h,l)}{\cos\theta_{i}}) \times \cos^{3}\theta_{i}  
\end{equation}
As it can be seen from the above mentioned expression it is an average value of $\exp(-\frac{a(h,l)}{\cos\theta}) \times \cos^{3}\theta$ for the angles $\theta \leq \Theta$. The estimation of (A.1) can be easily performed representing it in the integral form and assuming a uniform light emission:
\begin{equation}
S(a,\Theta) = \frac{\int_{0}^{2\pi}\int_{0}^{\Theta}\exp(-\frac{a(h,l)}{\cos\theta}) \times \cos^{3}\theta d\Omega}{\int_{0}^{2\pi}\int_{0}^{\Theta} d\Omega} =\frac{\int_{0}^{\Theta}\exp(-\frac{a(h,l)}{\cos\theta}) \times \cos^{3}\theta d\cos\theta}{\int_{0}^{\Theta} d\cos\theta}
\end{equation}
To perform the integration of the numerator of the relation (A.2) it is convenient to make a variable change like $x \equiv \cos\theta$ and  make the expressions shorter for $a(h,l)$ we will use the notation $a$:

\begin{equation}
f(x,a) = \int \exp(-\frac{a}{x}) \times x^{3} dx,
\end{equation}
which after integration by parts\cite{Wolfram} and using the relation for the exponential integral $\operatorname{E_i}\left(-x\right)=-\operatorname{E_1}\left(x\right)$  for $f(x,a)$ can be obtained:

\begin{equation}
f(x,a) = \frac{\mathrm{e}^{-\frac{a}{x}}\left(a^4\operatorname{E_1}\left(\frac{a}{x}\right)\mathrm{e}^\frac{a}{x}+6x^4-2ax^3+a^2x^2-a^3x\right)}{24}+const
\end{equation}

where $\operatorname{E_1}\left(\frac{a}{x}\right)$
is the well known exponential integral defined as $\operatorname{E_1}(x) = \int_{x}^{\infty} \frac{\exp(-u)}{u}du$ and has series representation:
\begin{equation}
\operatorname{E_1}(u) = -\gamma-\ln(u) - {\mathlarger{‎‎\sum}}_{k=1}^{\infty‎}\frac{(-1)^{k}\times u^{k}}{k\times k!}
\end{equation}

where $\gamma \approx 0.57721$ is the Euler constant. To minimize the computer precision effects for $u>1$ values is better the infinite sum in (A.5) evaluate as it is shown below:

\begin{equation}
{\mathlarger{‎‎\sum}}_{k=1}^{\infty‎}\frac{(-1)^{k}\times u^{k}}{k\times k!}={\mathlarger{‎‎\sum}}_{k=1}^{\infty‎}\frac{1}{k}{\mathlarger{\prod}}_{i=1}^{k‎} \frac{-u}{i}
\end{equation}
With this expression for the sum, 100 terms are sufficient to get stable evaluation for $u<20$ values. For $u>20$ probably the double precision is not sufficient for the stable evaluation of $E_{1}$ using the expressions (A.5 and A.6). For this work it is not so important, because the fibre length and the scintillator thickness in real detector designs can not be much larger than the attenuation lengths(always $u < 5$).

So for $S(a,\Theta)$ in (A.2) we will obtain:

\begin{equation}
S(a,\Theta) =\frac{f(1,a)-f(\cos\Theta,a)}{1-\cos\Theta}
\end{equation}
So the integral in the expression (5) can be written:

\begin{equation}
\int_{0}^{H}S(a,\Theta)dh =\lambda_{sc}\int_{a_{0}}^{a_{1}}S(a,\Theta)da=\lambda_{sc}\int_{a_{0}}^{a_{1}} \frac{f(1,a)-f(\cos\Theta,a)}{1-\cos\Theta}da
\end{equation}
where $a_{0}=a(0,l)$ is the fibre longitude in attenuation longitude units and $a_{1}=a(H,l)$ is the sum of scintillator thickness and fibre longitude also in attenuation longitude units as it has already been defined in the main text (see Eq. 6).
Ignoring attenuation in the scintillator, which means $S(a,\Theta)$ is independent from $h$ then for $\Phi_{dp}(\cos\Theta,a_{0},a_{0})$ we obtain:

\begin{equation}
\Phi_{dp}(\cos\Theta,a_{0},a_{0}) = H \times S(a,\Theta)
\end{equation}

Here before evaluating the integral mentioned above, we evaluate the indefinite integral:

\begin{equation}
\phi(x,a) = \int f(x,a)da
\end{equation}

Using expression (A.4) for $f(x,a)$ then for the $\phi(x,a)$ we obtain:

\begin{eqnarray}
\nonumber \phi(x,a) = \frac{x^5\mathrm{e}^{-\frac{a}{x}}}{24}\left((\frac{a}{x})^3+2(\frac{a}{x})^2+6(\frac{a}{x})\right) \\
-\frac{a^5}{24}\left(\frac{\gamma}{5}+\frac{5\ln \frac{a}{x}-1}{25} +{\mathlarger{‎‎\sum}}_{k=1}^{\infty‎}\frac{1}{k\times(k+5)}{\mathlarger{\prod}}_{i=1}^{k‎}\frac{-a}{x\times i}\right)+const
\end{eqnarray}

Finally for the $\Phi_{dp}(\cos\Theta,a_{0},a_{1})$ using the expression (A.7) we obtain: 

\begin{equation}
\Phi_{dp}(\cos\Theta,a_{0},a_{1}) =\lambda_{sc} \frac{(\phi(1,a_{1})-\phi(1,a_{0})-
(\phi(\cos\Theta,a_{1})-\phi(\cos\Theta,a_{0}) }{1-\cos\Theta}
\end{equation}

 Excluding the light attenuation in the expression (A.3), (ie a=0 in A4) for $\Phi(\cos\Theta)$ we will obtain:

\begin{equation}
\Phi_{dp}(\cos\Theta) = H\times\frac{1-cos^{4}\Theta}{4\times(1-cos\Theta)}
\end{equation}

The usage of a specular reflection with the reflection efficiency $\varepsilon_{sp}$ for the plane in front of the fibre matrix is similar to the one for the direct photons. So for the function $\Phi_{sp}(\cos\Theta,a_{1},a_{2})$ can be used the expression (A.10) with $a_{1}=a(H,l)$ and $a_{2}=a(2H,l)$

\begin{equation}
\Phi_{sp}(\cos\Theta,a_{1},a_{2}) =\varepsilon_{sp} \times\Phi_{dp}(\cos\Theta,a_{0},a_{1})
\end{equation}

  If the plain of the matrix also has specular reflectivity then we should sum all of reflections to obtain total light collection efficiency. For the matrix plane except reflection efficiency we have light loose connected with fibres. The efficiency connected with this lost $\varepsilon_{fm} = 1-S_{fib}/S_{tot}$ where $S_{fib}$ is the sum of the fibres' surface and $S_{tot}$ is the total matrix surface. So for the double reflection (at the beginning from the matrix plane and then from the plane in front of the fibres) for $\Phi_{spm}(\cos\Theta,a_{2},a_{3})$ we obtain an expression in a similar way as the previous one:
  
\begin{equation}
\Phi_{spm}(\cos\Theta,a_{2},a_{3}) =\varepsilon_{sp}^{2}\varepsilon_{fm} \times\Phi_{dp}(\cos\Theta,a_{0},a_{1}),
\end{equation}  
  
were $a_{3}=a(3H,l)$. So in this way we can estimate any time of reflections and then summing them we can have total contributions from the specular reflections. The number of the reflections is limited because of the value of $\varepsilon_{sp} \leq 0.8$ as well as $\varepsilon_{fm} \leq 0.95$.

\section{Efficiency estimation in case of diffuse reflection}\label{Efficiency estimation in case of diffuse reflection}

To estimate the integral in the expression (9), which is proportional to the amount of the light that reaches the reflective surface, at the beginning it should be evaluated by $\cos\theta_{1}$ and then by $h_{1}$. 

\begin{eqnarray}
L(\Psi,h_{1})=\int_{0}^{\Psi} \exp(-\frac{b(h_{1})}{\cos\theta_{1}})d\cos\theta_{1} 
\end{eqnarray}

This integral can be evaluated again using an indefinite integral. For the simplification we can substitute $x \equiv \cos\theta_{1}$ and again use $\operatorname{E_i}\left(-x\right)=-\operatorname{E_1}\left(x\right) $

\begin{eqnarray}
l(x,h_{1})=\int \exp(-\frac{b(h_{1})}{x})dx=x\exp(-\frac{b(h_{1})}{x})- b(h_{1})\operatorname{E_1}\left(\frac{b(h_{1})}{x}\right)+C
\end{eqnarray}

where $\operatorname{E_1}\left(\frac{a}{x}\right)$
is the well known exponential integral(see Appendix A). So the integral (B.1) can be estimated as:

\begin{eqnarray}
L(\Psi,h)=l(1,h_{1})-l(\cos\Psi,h_{1})
\end{eqnarray}

Performing the integration for $h_{1}$ of the expression (B.3) we obtain the estimation of the integral in (9): 
\begin{eqnarray}
I(\Psi,H)=\int_{0}^{H}l(1,h_{1})dh_{1}-\int_{0}^{H}l(\cos\Psi,h_{1})dh_{1}
\end{eqnarray}

Taking into account that $b(h_{1}) = h_{1}/\lambda_{sc}$, the expression (A5) for $\operatorname{E_1}\left(\frac{a}{x}\right)$ (see Appendix A) and again evaluating the indefinite integral $\int l(x,h_{1})dh_{1}$ we can estimate $I(\Psi,H)$.The evaluation of the indefinite integral $G(x,h)=\int l(x,h_{1})dh_{1}$ is shown below:

\begin{eqnarray}
\nonumber G(x,h_{1})= -\lambda_{sc}\times x^{2}\exp(-\frac{h_{1}}{\lambda_{sc}x}) \\
-\frac{h^{2}\times x}{\lambda_{sc}}\left(\frac{\gamma}{2}+\frac{2\ln \frac{h_{1}}{\lambda_{sc}x}-1}{4} +{\mathlarger{‎‎\sum}}_{k=1}^{\infty‎} \frac{1}{k\times(k+2)}{\mathlarger{\prod}}_{i=1}^{k‎}\frac{-h_{1}}{\lambda_{sc}\times x\times i}\right)+C
\end{eqnarray}

So for $I(\Psi,H)$ we obtain the expression below:

\begin{eqnarray}
I(\Psi,H)=G(1,H)-G(1,0)-G(\cos\Psi,H)+G(\cos\Psi,0)
\end{eqnarray}

 Without consideration of the light attenuation in the expression (B.1) for $I(\Psi,H)$ we will obtain:
 
\begin{eqnarray}
I(\Psi,H)=H\times(1-\cos\Psi)
\end{eqnarray}

To estimate the sum in the expression (10), which is shown below:
\begin{equation}
S_{dfr}(a_{1},\Theta) =  \frac{1}{N_{fH}} \times {\mathlarger{‎‎\sum}}_{i=0}^{N_{fH}‎} \exp(-\frac{a_{1}}{\cos\theta_{2i}}) \times \cos^{4}\theta_{2i}  
\end{equation}
As it can be seen from the expression above it is the average value of $\exp(-\frac{a(H,l)}{\cos\theta}) \times \cos^{3}\theta$ for the angles $\theta \leq \Theta$. So to estimate the sum we represent it in the integral form as it is done in Appendix A, assuming Lambert type distribution for the light reflection($\cos\theta$ low):
\begin{equation}
S_{dfr}(a_{1},\Theta) = \frac{\int_{0}^{2\pi}\int_{0}^{\Theta}\exp(-\frac{a_{1}}{\cos\theta}) \times \cos^{4}\theta d\Omega}{\int_{0}^{2\pi}\int_{0}^{\Theta} \cos\theta d\Omega} =\frac{\int_{0}^{\Theta}\exp(-\frac{a_{1}}{\cos\theta}) \times \cos^{4}\theta d\cos\theta}{\int_{0}^{\Theta}\cos\theta d\cos\theta}
\end{equation}
Denominator in the expression (B.9) is just for the normalization purpose.
To perform the integration of the numerator of the expression (B.9) it is convenient to make a variable change like $x \equiv \cos\theta$:

\begin{equation}
f_{dfr}(a_{1},x) = \int \exp(-\frac{a_{1}}{x}) \times x^{4} dx,
\end{equation}
which after the integration by parts\cite{Wolfram} can be obtained for $f_{dfr}(a_{1},x)$ :

\begin{equation}
f_{dfr}(a_{1},x) = - \frac{\mathrm{e}^{-\frac{a_{1}}{x}}\left(a_{1}^5 \operatorname{E_1}\left(\frac{a_{1}}{x}\right)\mathrm{e}^\frac{a_{1}}{x}-24x^5+6a_{1}x^4-2a_{1}^2x^3+a_{1}^3x^2-a_{1}^4x\right)}{120} + const
\end{equation}

where $\operatorname{E_1}\left(\frac{a_{1}}{x}\right)$
is the well known exponential integral(see A.5)  So for $S_{dfr}(a_{1},\Theta)$ in (B.9) we will obtain:

\begin{equation}
S_{dfr}(a_{1},\Theta) =\frac{f_{dfr}(a_{1},1)-f_{dfr}(a_{1},\cos\Theta)}{0.5\times (1-\cos^{2}\Theta)}
\end{equation}

Ignoring the light attenuation in the expression (B.11 ie $a_{1}=0$) for $S_{dfr}(0,\Theta)$ we will obtain:

\begin{equation}
S_{dfr}(0,\Theta) = \frac{2\times(1-cos^{5}\Theta)}{5\times(1-cos^{2}\Theta)}
\end{equation}

\section{Time spread estimation for the photon transport}\label{Time spread estimation for the photon transport}

The time spread estimation for $DP$ is based on the photon path spread estimation.  At the beginning it should be estimated the average path value and then the standard deviation using this average value for the photon path spread estimation. The average path value is estimated the same way as in the Appendix A. For each source point inside the scintillator the photon path length $L$ is calculated as:
\begin{equation}
L(l,h,\theta) = \frac{(l+h)}{\cos\theta}		
\end{equation}
where $l$ is the fibre length, $h$ is the distance of the source point from the fibre plane and $\theta$ is the angle between the vector connecting the light source to the given fibre and the vector that is perpendicular to the plane of the fibre matrix(see Fig a1).  The average value of the path length for the given fibre length $l$ and for the scintillator thickness $H$ is estimated with the standard procedure for the uniform angular photon emission:

\begin{equation}
\overline{L(l,h,\theta)} = \frac{\int_{0}^{H}\int_{0}^{2\pi}\int_{0}^{\Theta}\frac{(l+h)}{\cos\theta} d\Omega dh}{\int_{0}^{H}\int_{0}^{2\pi}\int_{0}^{\Theta} d\Omega} =\frac{\int_{0}^{H} \int_{1}^{\cos\Theta}\frac{(l+h)}{x} dx dh}{\int_{0}^{H}\int_{1}^{\cos\Theta} dx}
\end{equation}

where $\Omega$ is the solid angle $x=\cos\theta$ and integration is performed within fibre aperture angle $\Theta$. The Integration by the $h$ is performed for the interval from the zero to the scintillator thickness $H$. For the average value after integration we obtain the expression below:
\begin{equation}
\overline{L(l,H,\theta)} = \frac{(l+H/2)\times\ln\cos\Theta}{(\cos\Theta-1)} 
\end{equation}

In similar way can be estimated the standard deviation $\sigma_{L}$:

\begin{equation}
\frac{\sigma_{L}}{\overline{L}} =  \sqrt{\frac{(1+\frac{H}{l})\times (\cos\Theta-1)^{2}} {((1+\frac{H}{2l})^{2}\times\cos\Theta\times\ln^{2}\cos\Theta}-1}  
\end{equation}

 when $H/l \ll 1$ the expression (c4) depends only from the fibre aperture angle $\cos\Theta$ and the average path length $\overline{L}$.

\section{Figures}

 \begin{figure}[h]
  \centering
  \includegraphics[height=12cm]{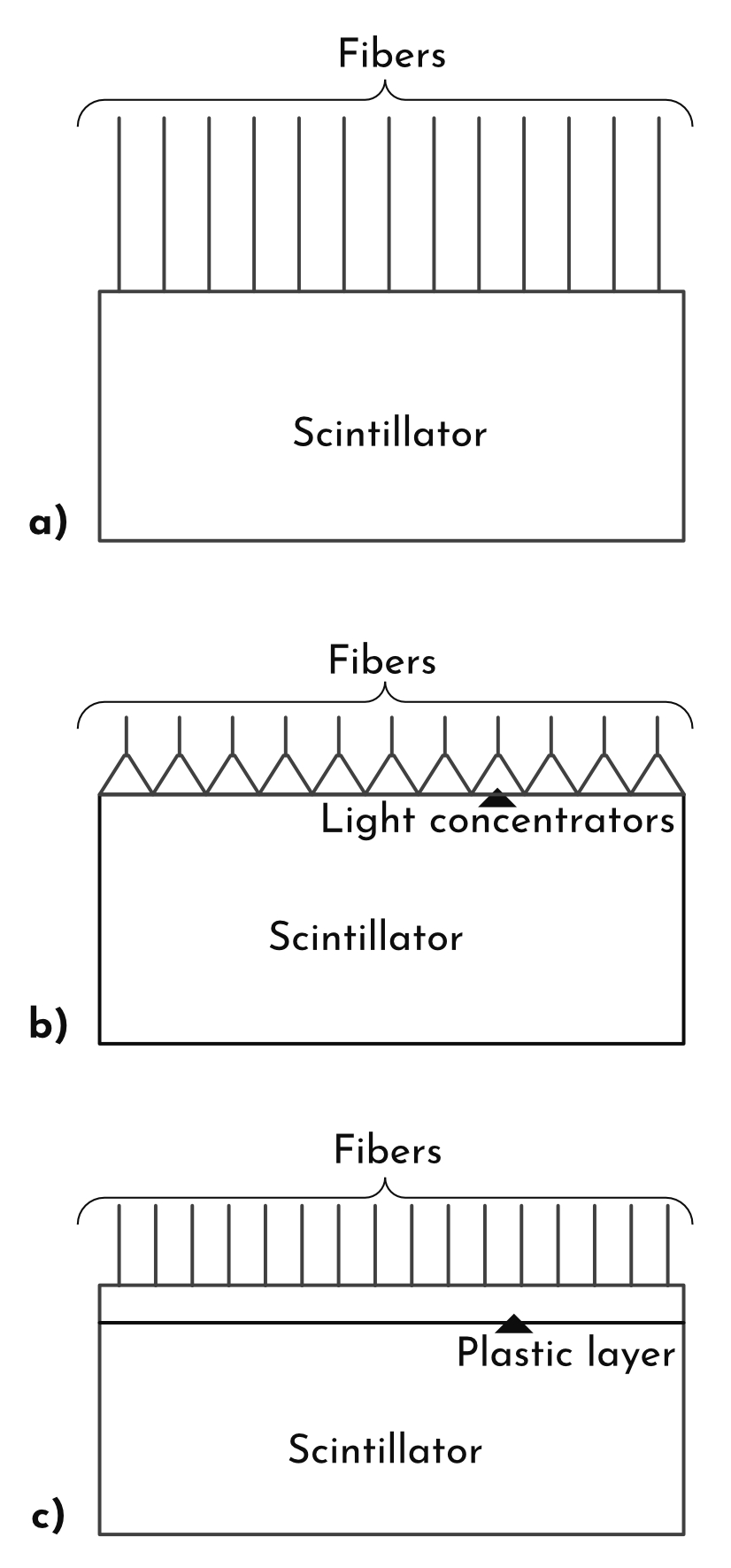}
  \caption{Three different light collection schemes.}
  \label{Fig.1}
 \end{figure}
 
 \begin{figure}[h]
  \centering
  \includegraphics[height=8cm]{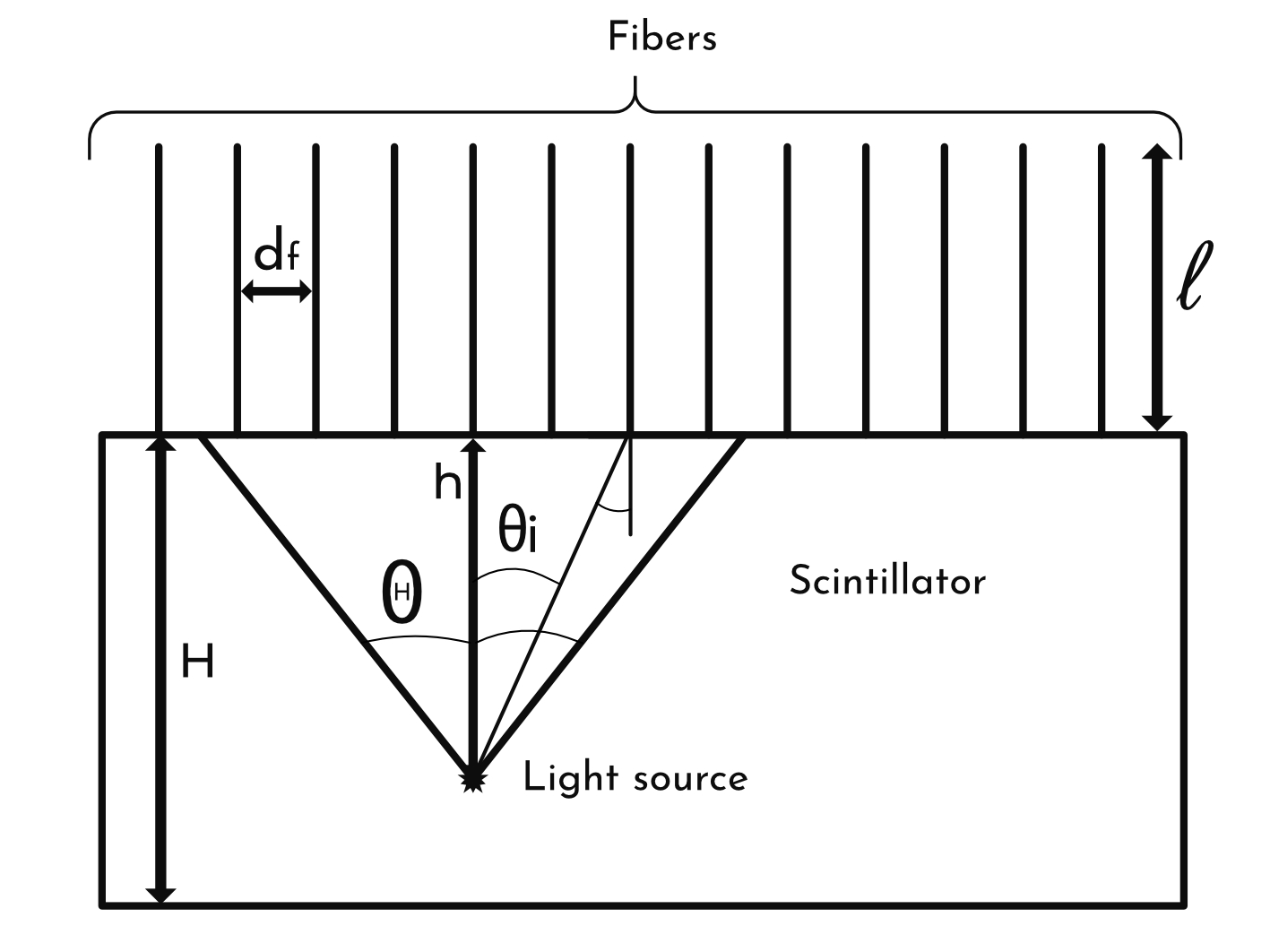}
  \caption{Light collection scheme from the point like source for direct photons. Description of all notations are in the text.}
  \label{Fig.2}
 \end{figure} 
 
 \begin{figure}[h]
  \centering
  \includegraphics[height=8cm]{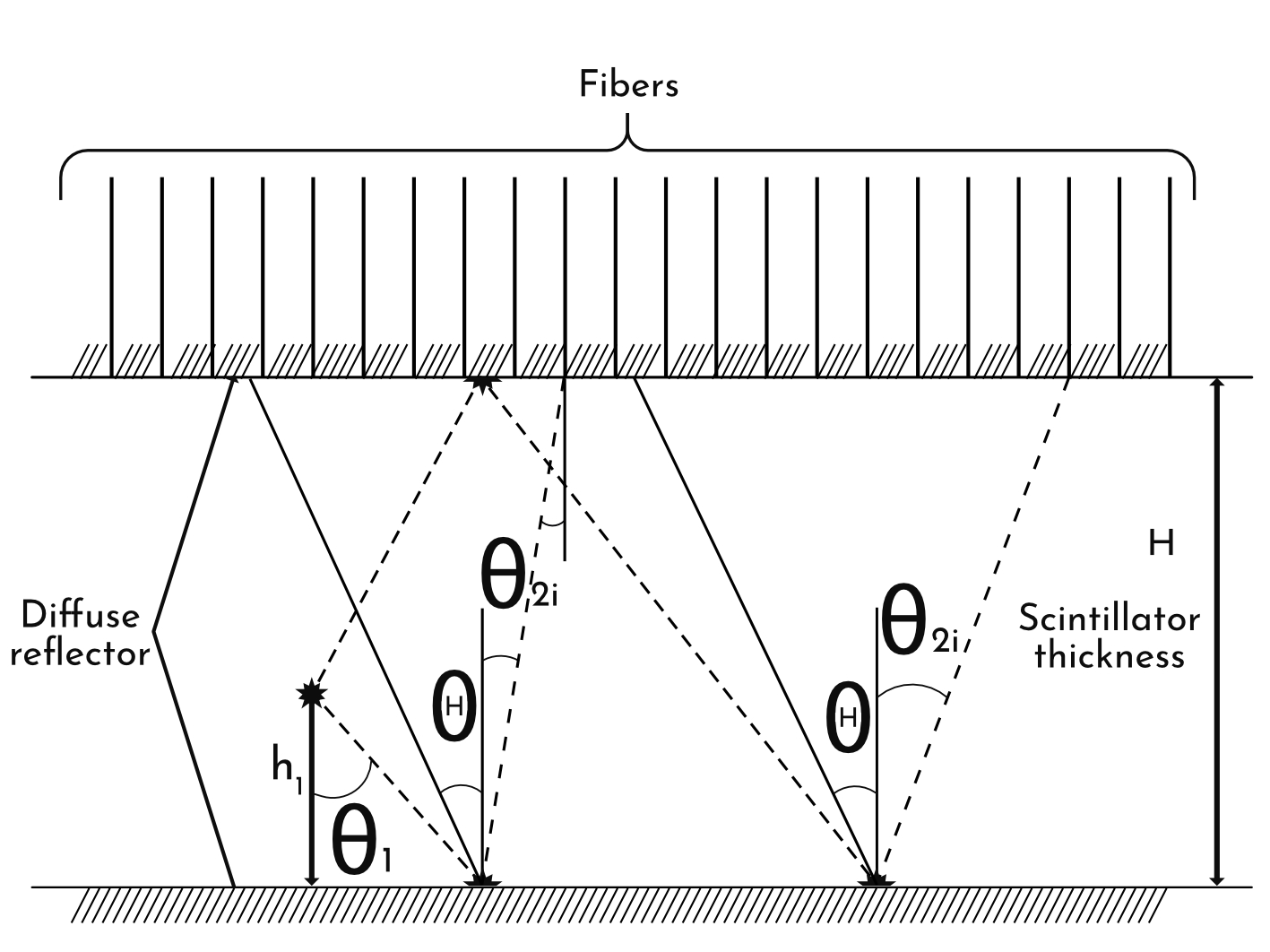}
  \caption{Light collection scheme from the point like source and for the  reflected photons. Description of all notations are in the text.}
  \label{Fig.3}
 \end{figure}  
 
 \begin{figure}[h]
  \centering
  \includegraphics[height=10cm]{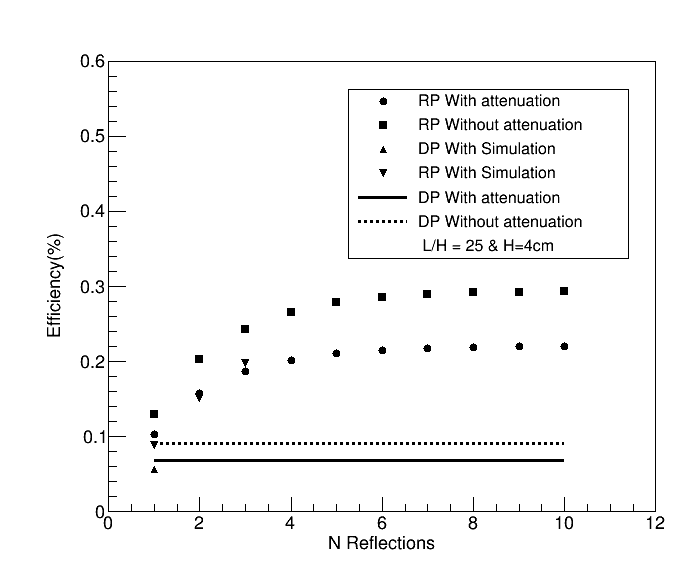}
  \caption{Light collection efficiency vs number of reflections. Lines show efficiency for DP with and without attenuation in the scintillator and in the fibres}
  \label{Fig.4}
 \end{figure}  
 
 \begin{figure}[h]
  \centering
  \includegraphics[height=10cm]{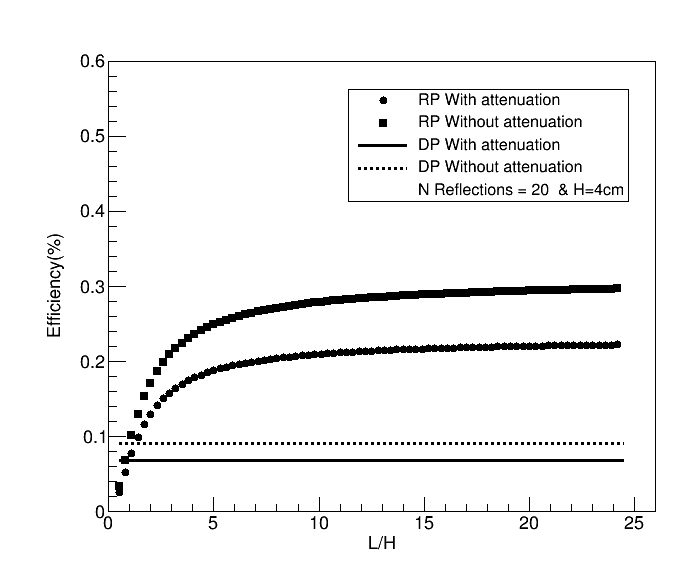}
  \caption{Light collection efficiency vs detector lateral sizes in the scintillator thickness units.}
  \label{Fig.5}
 \end{figure}   
 
 \begin{figure}[h]
  \centering
  \includegraphics[height=10cm]{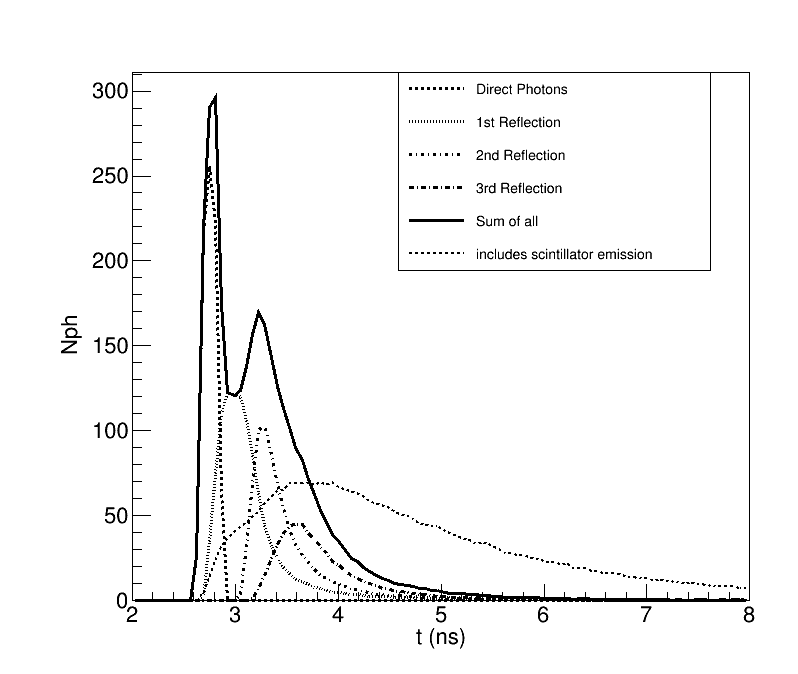}
  \caption{Pulse time wave forms for direct and reflected photons.}
  \label{Fig.6}
 \end{figure}  
  
\end{document}